\documentclass[]{jfm}

\usepackage{graphicx}
\usepackage{newtxtext}
\usepackage{newtxmath}
\usepackage{natbib}
\usepackage{hyperref}
\hypersetup{
    colorlinks = true,
    urlcolor   = blue,
    citecolor  = black,
}

\newcommand{\RomanNumeralCaps}[1]

\newcommand{\overbar}[1]{\mkern 1.5mu\overline{\mkern-1.5mu#1\mkern-1.5mu}\mkern 1.5mu}

\usepackage{booktabs}

\usepackage{bm}

\newcommand{\Ek}{{E\mkern -1.5mu k}}

\usepackage[svgnames]{xcolor}
\hypersetup{colorlinks=true, citecolor=NavyBlue, urlcolor=NavyBlue, linkcolor=NavyBlue}

\title{Discontinuous Transitions Towards Vortex Condensates in Buoyancy-Driven Rotating Turbulence}

\author{Xander M. de Wit\aff{1},
  Andr\'es J. Aguirre Guzm\'an\aff{1},
  Herman J. H. Clercx\aff{1}
  \and Rudie P. J. Kunnen\aff{1}
  \corresp{\email{r.p.j.kunnen@tue.nl}}}

\affiliation{\aff{1}Fluids and Flows group, Department of Applied Physics and J. M. Burgers Centre for Fluid Dynamics, Eindhoven University of Technology, P.O. Box 513, 5600 MB Eindhoven, Netherlands}

\begin{document}
\maketitle

\begin{abstract}
Using direct numerical simulations of rotating Rayleigh-B\'{e}nard convection, we explore the transitions between turbulent states from a 3D flow state towards a quasi-2D condensate known as the large-scale vortex (LSV). We vary the Rayleigh number $Ra$ as control parameter and study the system response (strength of the LSV) in terms of order parameters assessing the energetic content in the flow and the upscale energy flux. By sensitively probing the boundaries of the region of existence of the LSV in parameter space, we find discontinuous transitions and we identify the presence of a hysteresis loop as well as memoryless abrupt growth dynamics. We show furthermore that the creation of the condensate state coincides with a discontinuous transition of the energy transport into the largest mode of the system.
\end{abstract}

\begin{keywords}
\end{keywords}

\section{Introduction}
\label{sec:intro}

A hallmark feature of 3D turbulence is the forward energy cascade, transporting kinetic energy from large scales to ever smaller scales as described by the celebrated theory of \citet{Kolmogorov1941}. In many geophysical and astrophysical flows, however, velocity fluctuations are largely suppressed in one direction as a consequence of, for example, confinement \citep{Benavides2017,Musacchio2017,Musacchio2019}, strong magnetic fields \citep{Alexakis2011,Seshasayanan2014} or fast rotation \citep{Smith1996,Seshasayanan2018,Pestana2019,VanKan2020}, rendering the flow quasi-2D. This leads to the development of an inverse energy flux, akin to fully 2D turbulence \citep{Kraichnan1967,Batchelor1969}, transporting energy from smaller to larger scales. Ultimately, this can lead to accumulation of kinetic energy at the largest available scales, followed by condensation into a vertically coherent large-scale vortex (LSV) structure at the domain size \citep[see][for a recent review]{Alexakis2018}. These LSVs are believed to play a crucial role in for example the formation of the Earth's magnetic field \citep{Aurnou2015,Guervilly2015,Roberts2013}.

Following the framework that is brought forward in \citet{Alexakis2018}, we aim to classify the transition from a 3D forward cascading state to the condensate state by considering the behaviour of an order parameter that measures the strength of the LSV as a function of a control parameter of the flow throughout this transition. Then one can observe either a smooth transition, a continuous transition with discontinuous derivative, or a discontinuous transition. This categorisation of the transition into the condensate state has shown to be an insightful approach in various other quasi-2D flow systems \citep{Alexakis2015,Seshasayanan2018,VanKan2019,Yokoyama2017}.

These earlier works, however, have focused on more artificial, idealised flow models where the turbulent forcing occurs at a single well-defined length scale. Here, we characterise the LSV transition in a natural, broadband-forced system of rotating convection, which is ubiquitous in geophysical and astrophysical flows. In this system, \citet{Favier2019} have shown a bistability of an LSV with a non-LSV state at the same parameter values, depending on the initial conditions. The natural buoyant forcing over a broad range of scales obfuscates the strict separation of the injection, dissipation and condensation scale. Although one may expect that in such natural and vigorously fluctuating turbulent systems, any transitions between different states are washed out and become gradual, we find that the transition towards the condensate state is in fact sharp and discontinuous.

Such abrupt transitions between turbulent states in a more general sense are a remarkable feature of fluid turbulence and have received much recent interest, being observed in various different flow settings, e.g. in torque measurements of Taylor-Couette and Von Kármán flows \citep{Huisman2014,Ravelet2004,Saint-Michel2013}, in states of stochastically forced 2D and 3D turbulence \citep{Bouchet2009,Bouchet2019,Iyer2017}, and in reversals of large scale dynamics in thin layers \citep{Sommeria1986,Michel2016,Dallas2020}. These types of abrupt transitions are surmised to play an important role in, for example, climate research \citep{Weeks1997,Herbert2020,Jackson2018} and in understanding the geomagnetic reversal \citep{Petrelis2009,Berhanu2007}.

\begin{figure}
    \centering
    \includegraphics[width=0.7\linewidth]{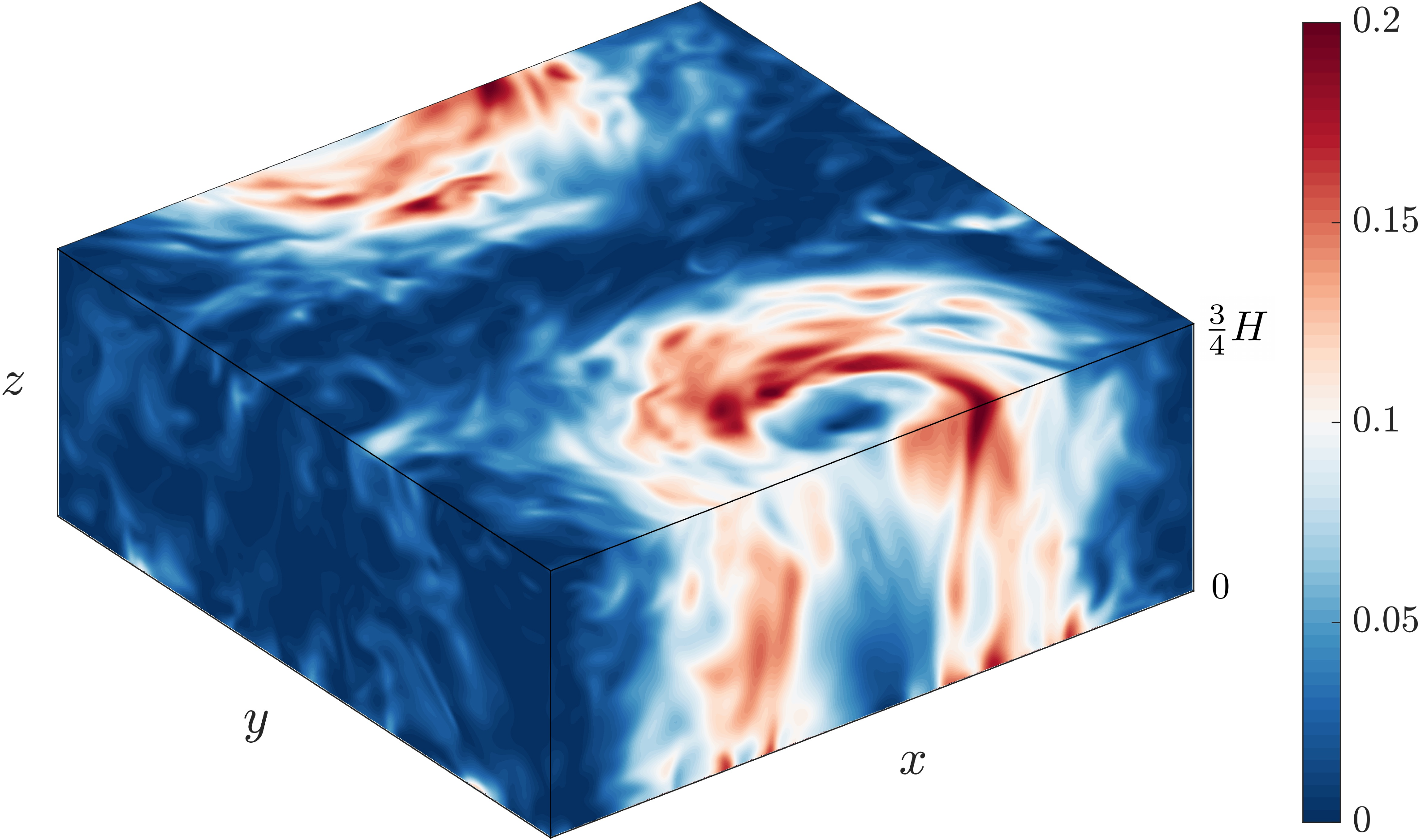}
    \caption{Snapshot of horizontal kinetic energy (in units of $U^2$) of the LSV-forming case $Ra=1.7\times10^7$, truncated at three-quarter height to reveal a cross section of the LSV.}
    \label{fig:lsv_example}
\end{figure}

\section{Numerical approach}\label{sec:numerical_approach}
We consider the canonical system of rotating Rayleigh-B\'{e}nard convection, in which the flow is driven by buoyancy through a temperature difference $\Delta T$ between the bottom and top of the domain, whilst being simultaneously affected by strong background rotation $\Omega$ along the vertical axis. The input space to this problem is described by three dimensionless numbers: the Rayleigh number $Ra=g\alpha\Delta T H^3/(\nu\kappa)$, quantifying the strength of the thermal forcing, the Ekman number $\Ek=\nu/(2\Omega H^2)$, representing the (inverse) strength of rotation and the Prandtl number $\Pran=\nu/\kappa$, containing the diffusive properties of the fluid. Here, $g$ denotes gravitational acceleration, $H$ is the domain height and $\alpha$, $\nu$ and $\kappa$ represent the thermal expansion coefficient, kinematic viscosity and thermal diffusivity of the fluid, respectively. The system is non-dimensionalized into convective units using $H$, $\Delta T$ and the free-fall velocity $U=\sqrt{g\alpha\Delta T H}$.

We solve the full governing set of Boussinesq Navier-Stokes equations through direct numerical simulation (DNS), employing the finite-difference code described in \citet{Verzicco1996} and \citet{Ostilla-Monico2015} on a Cartesian grid with periodic sidewalls and stress-free boundary conditions at the top and bottom. For the width $D$ of the domain, we choose $D/H=10\mathcal{L}_c$ with $\mathcal{L}_c=4.8\Ek^{1/3}$ the most unstable wavelength at onset of convection \citep{Chandrasekhar1961}. The complete set of input parameters as well as the employed resolutions are provided in table~\ref{tab:input}. Note that we include the convective Rossby number $Ro=U/(2\Omega H)=\Ek\sqrt{Ra/\Pran}$ for reference. A validation of the grid resolution is provided in appendix~\ref{appA}.

\begin{table}
    \renewcommand\arraystretch{1.1}
    \centering
    \caption{The different series of input parameters used in this work.}\label{tab:input}
    \makebox[\textwidth][c]{
    \begin{tabular}[t]{lccccccc}
        \toprule
        &\#runs&$Ra$&$\Ek$&$\Pran$&$Ro$&$D/H$&Resolution\\
        \midrule
        Low-$Ra$ transition&14&$[2\times10^6 : 1\times10^7]$&$10^{-4}$&1&$[0.14:0.32]$&2.24&$256\times256\times128$\\
        Intermediate $Ra$&2&$[1.3\times10^7:1.7\times10^7]$&$10^{-4}$&1&$[0.36:0.41]$&2.24&$256\times256\times136$\\
        High-$Ra$ transition&30&$[2\times10^7 : 5\times10^7]$&$10^{-4}$&1&$[0.45:0.71]$&2.24&$256\times256\times144$\\
        Ensemble&100&$6\times10^6$&$10^{-4}$&1&0.24&2.24&$128\times128\times72$\\
        \bottomrule
    \end{tabular}
    }
\end{table}

For numerical convenience, we use $\Pran=1$ and $\Ek=10^{-4}$, for which stable LSVs have been observed in earlier DNSs over a limited range of $Ra$ \citep{Favier2014,Guervilly2014,Favier2019}. Upon increasing $Ra$ from the onset of convection, the two boundaries of the region of existence of the LSV are crossed. At the low-$Ra$ transition, the LSV develops as sufficient turbulent forcing is obtained to set-up the upscale transport into the condensate, whereas at the high-$Ra$ transition, the LSV breaks down as too strong thermal forcing renders the flow insufficiently rotationally constrained, breaking the quasi-2D conditions for upscale energy flux \citep{Favier2014,Guervilly2014}. We carry out a total of 46 runs at varying $Ra$, concentrated around both LSV transitions.

In order to analyse the LSV, we decompose the flow field $\bm{u}=u\bm{e}_x+v\bm{e}_y+w\bm{e}_z$ into a 2D (vertically averaged) barotropic flow and a 3D (depth-dependent) baroclinic flow \citep[following][]{Guzman2020,Favier2019,Favier2014,Julien2012,Maffei2021,Rubio2014}, i.e. $\bm{u}=\bm{u}^\textrm{2D}+\bm{u}^\textrm{3D}$, where
\begin{equation}
    \bm{u}^\textrm{2D}=\overbar{u}\bm{e}_x+\overbar{v}\bm{e}_y,\quad \bm{u}^\textrm{3D}=\underbrace{(u-\overbar{u})}_{u'}\bm{e}_x+\underbrace{(v-\overbar{v})}_{v'}\bm{e}_y+w\bm{e}_z,
\end{equation}
where the overbar $\overbar{\vphantom{a}\textrm{...}}$ denotes vertical averaging. Since the LSV is a largely vertically coherent structure (see figure~\ref{fig:lsv_example}), it resides primarily in the 2D field, whereas the turbulent baroclinic fluctuations are captured by the 3D field. Accordingly, we decompose the total kinetic energy $E_\textrm{tot}=\tfrac{1}{2}\langle|\bm{u}|^2\rangle$ into 2D and (horizontal and vertical) 3D contributions $E_\textrm{tot}=E^\textrm{2D}+E_H^\textrm{3D}+E_V^\textrm{3D}$ as
\begin{equation}
    E^\textrm{2D}=\tfrac{1}{2}\langle\overbar{u}^2+\overbar{v}^2\rangle,\quad E_H^\textrm{3D}=\tfrac{1}{2}\langle u'^2+v'^2\rangle,\quad E_V^\textrm{3D}=\tfrac{1}{2}\langle w^2\rangle,
\end{equation}
where angular brackets $\langle ... \rangle$ represent an average over the full spatial domain. We also consider the energy spectrum of the 2D flow from its Fourier transform $\hat{\bm{u}}^\textrm{2D}_{k_x k_y}$ as
\begin{equation}
    \hat{E}^\textrm{2D}(K)=\sum_{K\leq\sqrt{k_x^2+k_y^2}<K+1} \tfrac{1}{2}|\hat{\bm{u}}^\textrm{2D}_{k_x k_y}|^2,
\end{equation}
where we normalise the wavenumber $K$ by the box-size mode $2\pi/D$, such that the LSV occupies the $K=1$ mode of the spectrum.


\section {Results and discussion}\label{sec:results}
\begin{figure}
    \centering
    \includegraphics[width=0.55\linewidth]{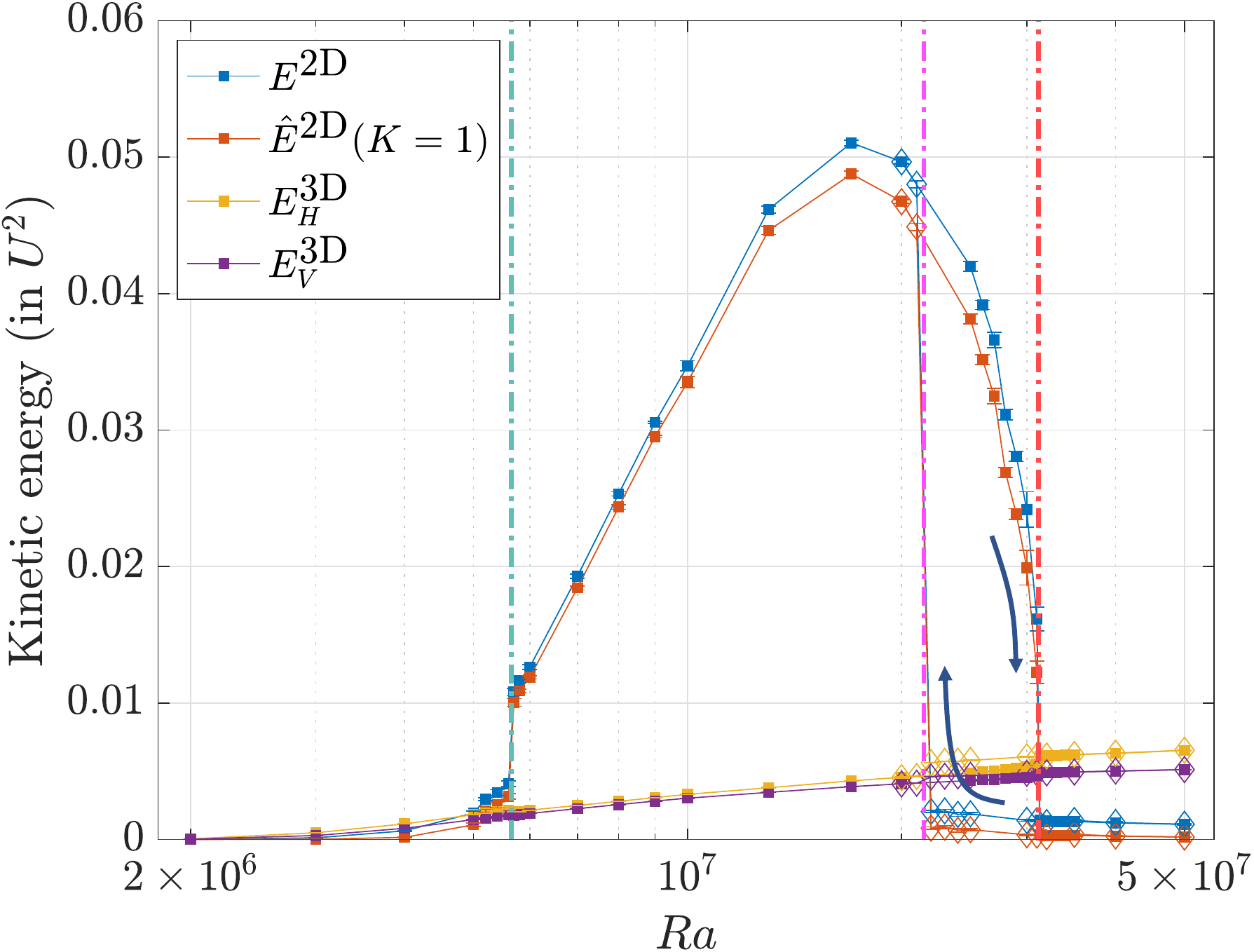}
    \caption{Averaged kinetic energy components as a function of $Ra$. Filled squares: upper hysteretic branch of the high-Ra transition; open diamonds: lower hysteretic branch. The cyan dashed-dotted line denotes the low-$Ra$ LSV transition, whilst the magenta and red lines denote the LSV transition of the lower and upper branch of the high-$Ra$ transition, respectively.}
    \label{fig:kins}
\end{figure}

In figure~\ref{fig:kins}, the different components of kinetic energy are provided over the range of considered $Ra$, crossing both LSV transitions. From the (largest mode of) 2D energy that captures the LSV, we find a substantial discontinuity at both boundaries of the LSV state. At the high-$Ra$ transition, we find that this transition is hysteretic: the morphology of the flow depends on its history c.q. its initial conditions. These findings are in line with \citet{Favier2019}, showing this bistability of an LSV and a non-LSV state for one case in this parameter range. To study this hysteresis loop, we initialise simulations using flow snapshots from a preceding $Ra$. For decreasing $Ra$ from a non-LSV state, the lower branch of the hysteresis loop is followed (open diamonds), whilst for increasing $Ra$ from an LSV state, the flow adheres to the upper hysteretic branch (filled squares), see figure~\ref{fig:kins}. Note the remarkably large discontinuity in the lower hysteretic branch, where the flow transitions directly from a non-LSV state into nearly the strongest LSV forming state. Hysteresis in the LSV transition has also been observed in a rotating flow system with a \emph{sharp bandwidth} Taylor-Green forcing \citep{Yokoyama2017}.

\begin{figure}
    \includegraphics[width=\textwidth]{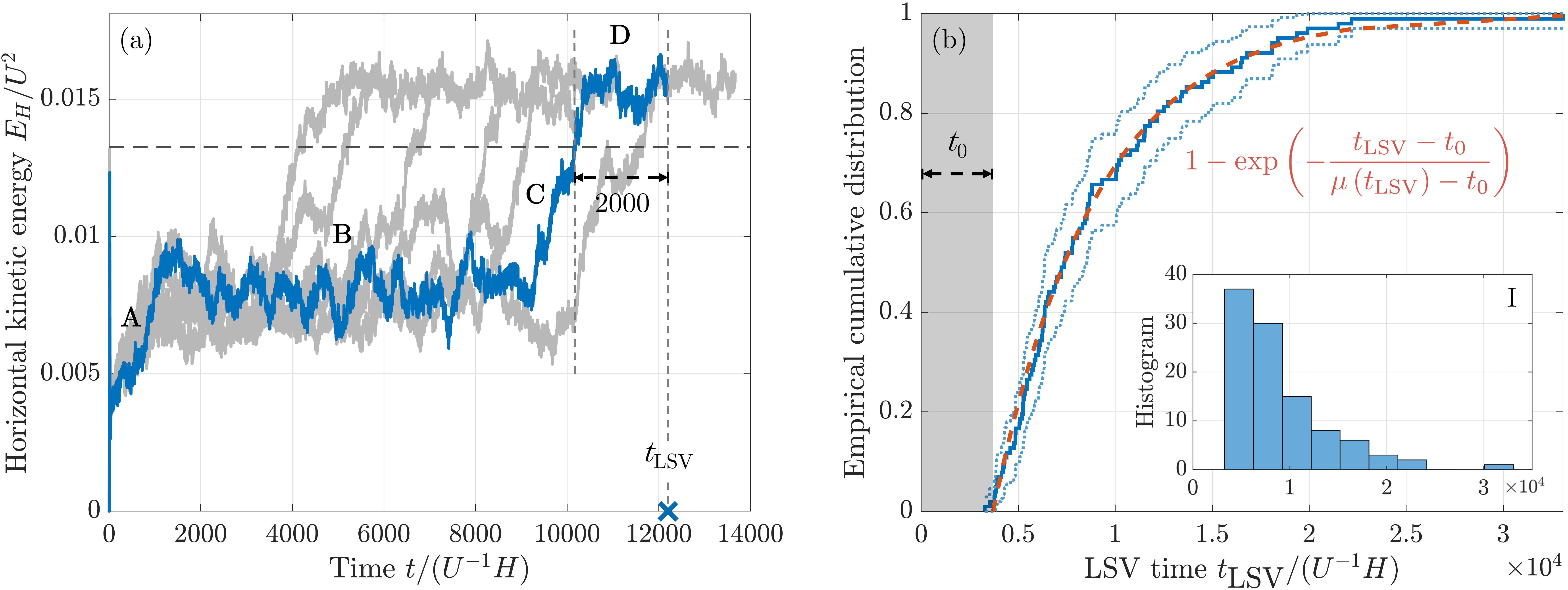}
    \caption{Panel (a): time evolution of horizontal kinetic energy of seven of the runs in the ensemble of the LSV-forming case $Ra=6\times 10^6$ close to the transition. We distinguish four stages: an initialisation phase (A), a metastable underdeveloped plateau state showing alternating jets (B), (quick) growth of the LSV (C) and the stably developed LSV state (D). The horizontal dashed line denotes the threshold to be crossed and sustained for 2000 convective time units ($U^{-1}H$) before the LSV is said to be completely developed, which defines the time point $t_\textrm{LSV}$, depicted by the blue cross. Panel (b): cumulative distribution of $t_\textrm{LSV}$ of the ensemble (blue solid line) with 95\% confidence bounds (blue dotted lines). It is fitted by an exponential distribution (red dashed line) as provided by (\ref{eq:fit_dist}) and in the figure. Inset (I) shows the histogram corresponding to the same distribution.}
    \label{fig:barrier}
\end{figure}

At the low-$Ra$ transition, on the other hand, no hysteresis is observed (using increments in $Ra$ of $\sim$2\%). Considering the cases directly above the LSV transition, however, we find that the growth of the LSV from a non-LSV state is non-monotonic: the flow shows an evident plateau during which the LSV has not yet developed, before finally growing relatively suddenly into the stable LSV state, see figure~\ref{fig:barrier}a. The flow in this plateau state shows morphological similarities with the jet state observed in rectangular domains \citep{Guervilly2017,Julien2018}, but alternates between being predominantly in the $x$- and $y$-direction. We hypothesise that the peculiar growth behaviour found here signifies a memoryless abrupt growth process, much akin to the nucleation \& growth type of dynamics that is observed in plentiful different systems throughout physics \citep{Matsumoto2002,Watanabe2010,Garmann2019,Metaxas2019}. To substantiate this conjecture, we simulate an ensemble of 100 additional runs at $Ra=6\times10^6$ with statistically perturbed initial conditions, using a reduced resolution of $128\times128\times72$ for computational feasibility. The hypothesised abrupt memoryless growth would then predict an exponential distribution of the waiting time spent in the metastable plateau state (stage B in figure~\ref{fig:barrier}a).

To investigate the distribution of these waiting times, we define a time point $t_\textrm{LSV}$ at which the LSV is said to have stably developed once a threshold of horizontal kinetic energy is surpassed and sustained for 2000 convective time units, see figure~\ref{fig:barrier}a. The obtained empirical cumulative distribution is then fitted with an exponential distribution
\begin{equation}\label{eq:fit_dist}
    CDF(t_\textrm{LSV})=1-\exp\bigg(-\frac{t_\textrm{LSV}-t_0}{\underbrace{\mu(t_\textrm{LSV})-t_0}_{\tau_W}}\bigg),
\end{equation}
where the fit parameter $t_0$ can be interpreted as the (fixed) contributions of the initialisation, growth and stable LSV stages (A, C and D in figure~\ref{fig:barrier}a). Here, $\mu(t_\textrm{LSV})$ denotes the mean of $t_\textrm{LSV}$, providing the maximum-likelihood estimate for the typical waiting time $\tau_W$ in this distribution. Figure~\ref{fig:barrier}b shows that there is excellent agreement between the hypothesised and the obtained distribution: the exponential distribution remains everywhere in between the 95\% confidence bounds of the empirical distribution. Signs of exponentiality of waiting times have been observed in the LSV-forming system of \emph{sharp bandwidth} forced thin-layer turbulence \citep{VanKan2019a}. Our findings indicate that, indeed, turbulent fluctuations randomly trigger the growth of the LSV, giving rise to this memoryless abrupt growth dynamics. Similar sudden growth behaviour is also found near the ends of the hysteresis loop in the high-$Ra$ transition, although an extended analysis of how the mean transition time evolves for changing $Ra$, repeating the ensemble average approach for all points in the considered parameter space, is currently out of computational reach for rotating convection.

Considering the observed hysteretic behaviour as well as the exponentially distributed waiting times, an analogy of this transition with first-order phase transitions in equilibrium statistical mechanics seems befitting \citep{Binder1987}. However, although it is conjectured that in particular the relatively more weakly dissipative large scales of the flow may be in resemblance with thermal equilibrium states \citep{Alexakis2018,Bouchet2012}, ultimately, the chaotic and dissipative nature of turbulence makes the analogy with equilibrium statistical mechanics indirect. A more immediate interpretation of the transition in this fluctuating dynamical system would be in terms of non-linear bifurcations. Then, the condensate transition as observed here can be interpreted as a subcritical non-linear bifurcation, giving rise to two distinct attractors (indeed, the LSV state and the non-LSV state) which remain separated in phase space. Such noise induced transitions between attractors are also known to be memoryless, yielding exponentially distributed waiting times \citep{Kraut1999}.

To understand how the LSV is energetically sustained, we compute the mode-to-mode kinetic energy transfer \citep[see][]{Dar2001,Alexakis2005,Mininni2005,Mininni2009,Verma2017,Verma2019}, distinguishing the 3D to 2D (baroclinic to barotropic) transport \citep[following][]{Rubio2014,Guzman2020}
\begin{subequations}
\begin{equation}
    T_\textrm{3D}(K,Q)=-\Big\langle\bm{u}^\textrm{2D}_K\cdot(\overbar{\bm{u}^\textrm{3D} \cdot \boldsymbol{\nabla} \bm{u}^\textrm{3D}_Q})\Big\rangle,
\end{equation}
and the 2D to 2D (barotropic to barotropic) transport
\begin{equation}
    T_\textrm{2D}(K,Q)=-\Big\langle\bm{u}^\textrm{2D}_K\cdot(\bm{u}^\textrm{2D}\cdot\boldsymbol{\nabla}\bm{u}^\textrm{2D}_Q)\Big\rangle,
\end{equation}
\end{subequations}
describing the energetic transport into the Fourier-filtered 2D flow field $\bm{u}^\textrm{2D}_K$ of mode $K$ from 3D and 2D modes $Q$ through triadic interactions arising from the advective term of Navier-Stokes. If $T_\textrm{3D},T_\textrm{2D} > 0$, there is a net transfer of kinetic energy from mode $Q$ to mode $K$ and vice-versa. We also consider the transport into 3D mode $K$ from the full (unfiltered) flow components $\mathcal{T}_\textrm{3D}(K)=\sum_Q T_\textrm{3D}(K,Q)$ and $\mathcal{T}_\textrm{2D}(K)=\sum_Q T_\textrm{2D}(K,Q)$, by summing over the donating modes $Q$.

\begin{figure}
    \centering
    \includegraphics[width=0.7\linewidth]{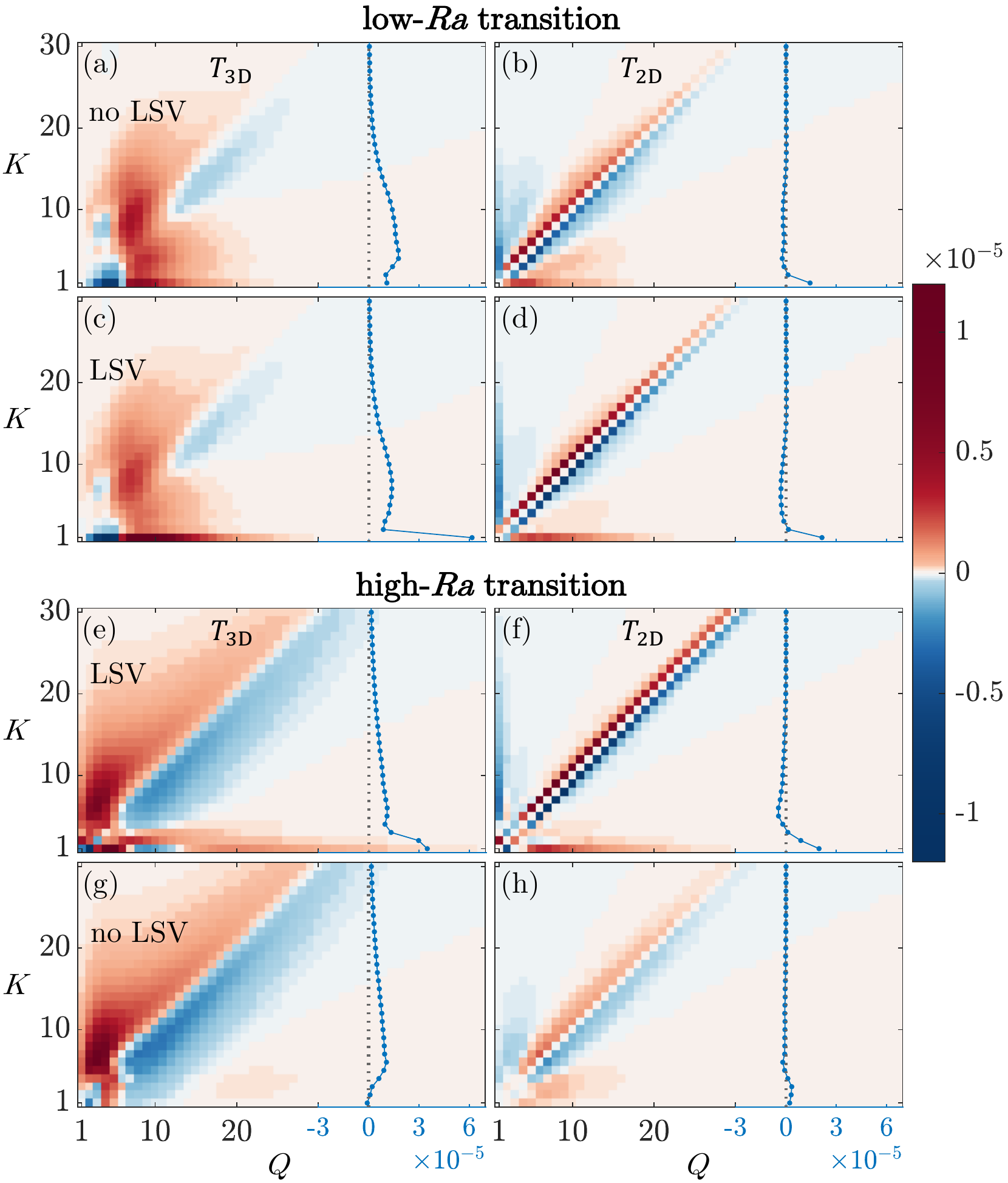}
    \caption{Time-averaged kinetic energy transport from 3D (left panels) or 2D (right panels) modes $Q$ to 2D modes $K$, i.e. $T_\textrm{3D}(K,Q)$ and $T_\textrm{2D}(K,Q)$, respectively, in units of $U^3H^{-1}$. Blue lines denote $\mathcal{T}_\textrm{3D}(K)$ (left) and $\mathcal{T}_\textrm{2D}(K)$ (right). The low-$Ra$ transition is crossed from (a,b) [$Ra=5.6\times10^6$] to (c,d) [$Ra=5.7\times10^6$], whilst the high-$Ra$ transition of the upper hysteretic branch is crossed from (e,f) [$Ra=3.10\times10^7$] to (g,h) [$Ra=3.13\times10^7$], as also depicted in figure~\ref{fig:trans}.}
    \label{fig:transfer_maps}
\end{figure}

The results for the shell-to-shell energy transfer throughout the LSV transitions are provided in figure~\ref{fig:transfer_maps}. The two main energy fluxes are apparent in both $T_\textrm{3D}$ and $T_\textrm{2D}$. Near the diagonal, one can observe the direct forward cascade, transporting energy from $Q$ to slightly higher modes $K$. Note here, that while the $T_\textrm{2D}$ self-interaction must be symmetric by definition $T_\textrm{2D}(K,Q)=-T_\textrm{2D}(Q,K)$, this does not apply to $T_\textrm{3D}$ as it describes the energetic interactions between scales of the 2D component and 3D component of the flow. In the bottom row $K=1$, on the other hand, the upscale energy flux into the LSV can be appreciated. This energy flux is non-local: energy is transported directly from virtually all scales in the system into the box scale of the LSV, without participation of intermediate scales.

\begin{figure}
    \centering
    \includegraphics[width=0.55\linewidth]{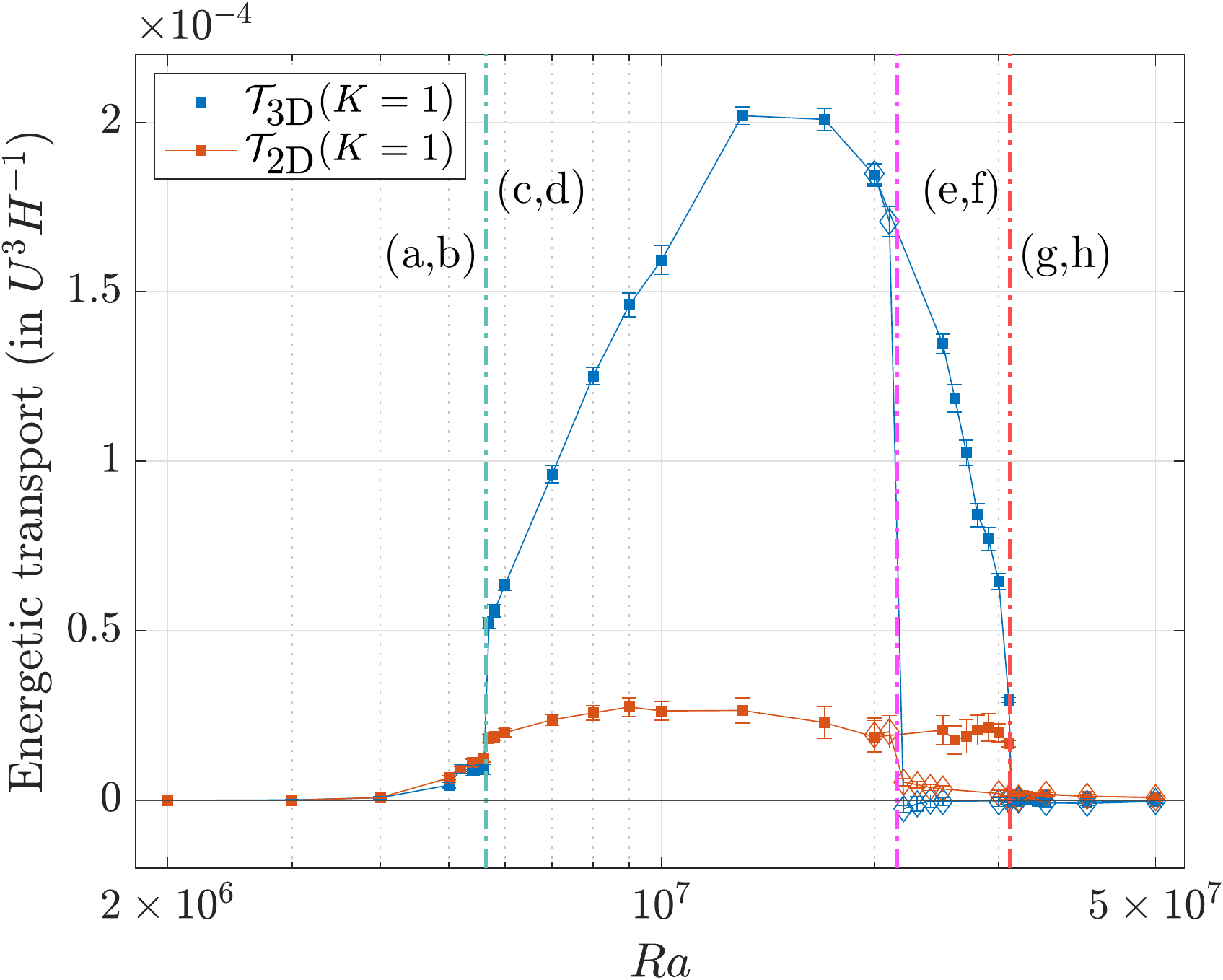}
    \caption{Total 3D $\mathcal{T}_\textrm{3D}(K=1)$ (blue) and 2D $\mathcal{T}_\textrm{2D}(K=1)$ (red) transport of kinetic energy into the 2D $K=1$ mode averaged over time as a function of $Ra$. Symbols and vertical lines are as in figure~\ref{fig:kins}. Labels (a-h) depict the corresponding transfer maps in figure~\ref{fig:transfer_maps}.}
    \label{fig:trans}
\end{figure}

Figure~\ref{fig:trans} shows the energetic transport into the box-size mode as a function of $Ra$. Note that this considers the transport from the full, unfiltered 3D and 2D flow components into the LSV, that is, a sum over the donating scales in the bottom row $K=1$ of the transfer maps in figure~\ref{fig:transfer_maps}. It makes clear that also the upscale transport into the LSV exhibits an evident discontinuous transition, both in $\mathcal{T}_\textrm{3D}(K=1)$ as well as, albeit to a lesser degree, in $\mathcal{T}_\textrm{2D}(K=1)$. Importantly, the figure indicates that it is the 3D transport that is the dominant component in the forcing of the LSV.


We argue that this upscale transport provides a clue to understand the physical mechanism behind the observed sudden growth and hysteretic behaviour. As also detailed in \citet{Rubio2014} and \citet{Favier2019}, the upscale transport contains a positive feedback loop, where the presence of the LSV itself enhances the upscale transport into the box-size mode. This agrees with our observation that the energetic transport into the LSV increases discontinuously as the LSV is created. The exact nature of how the LSV interacts energetically with its 3D turbulent background is an interesting non-trivial question to explore in future work, which goes beyond considerations purely in Fourier space as well, by looking at individual vortex interactions, for example. The existence of such a positive feedback loop, however, seems intuitive: the predominantly cyclonic LSV locally increases the total vorticity (background rotation + flow vorticity), thereby strengthening the quasi-2D conditions and hence the upscale transport. This mechanism allows the LSV to develop once its growth is triggered by rare turbulent fluctuations and lets the LSV remain stably self-sustained over the hysteresis loop.

Since we consider specifically a cross section of the full parameter space for varying $Ra$, the influence of other parameters on the transitional dynamics, such as $\Ek$, $\Pran$ and also the aspect ratio $D/H$ (since the domain width is the principal dynamical scale of the LSV), remains an open question. The morphology of the LSV in the asymptotically reduced model for $\Ek\to0$ is studied in its total parameter space in more detail in \citet{Maffei2021}. For the transition specifically, however, one can argue that as both attractors are expected to shift in continuous fashion through the phase space, only quantitative changes to the observed transitional dynamics are expected as the other control parameters are varied in vicinity to the computationally tractable values considered here. Nonetheless, the possibility that the discontinuity in the transition vanishes in a certain limit if both attractors would shift to coincide cannot be ruled out from the current simulations; the asymptotically reduced model seems appropriate to investigate this premise for the limit $\Ek\to0$.

\section{Conclusions}\label{sec:conclusions}
We have described the fluid turbulence transition into a quasi-2D condensate state in a natural broadband-forced system of rotating Rayleigh-B\'{e}nard convection, where the transition is sharply discontinuous, in spite of the lack of a clear separation of scales. We provide evidence of memoryless abrupt growth dynamics and hysteresis in these transitions, raising the picture of a double attractor phase space with a subcritical noise-induced transition between them. Furthermore, the correspondence of our findings with certain aspects of the LSV transition in other, artificially forced, flow systems, as remarked in the text, ultimately shows that this peculiar type of transition is a relevant and robust phenomenon that is expected to survive even in the geo- and astrophysically relevant flow of rotating convection, being one of the principal sources of fluid motion in nature.

\section*{Acknowledgments}
We thank W.G. Ellenbroek for careful reading of the manuscript. A.J.A.G. and R.P.J.K. received funding from the European Research Council (ERC) under the European Union’s Horizon 2020 research and innovation programme (Grant Agreement No. 678634). We are grateful for the support of the Netherlands Organisation for Scientific Research (NWO) for the use of supercomputer facilities (Cartesius) under Grants No. 2019.005 and No. 2020.009.

\section*{Declaration of Interests}
The authors report no conflict of interest.

\appendix
\section{Grid validation}\label{appA}
We employ a Cartesian grid that is uniform in the $x$- and $y$-direction, but non-uniform in the $z$-direction, clustering grid cells more closely near the top and bottom of the domain to properly resolve the boundary layers.

\begin{figure}
    \centering
    \includegraphics[width=\textwidth]{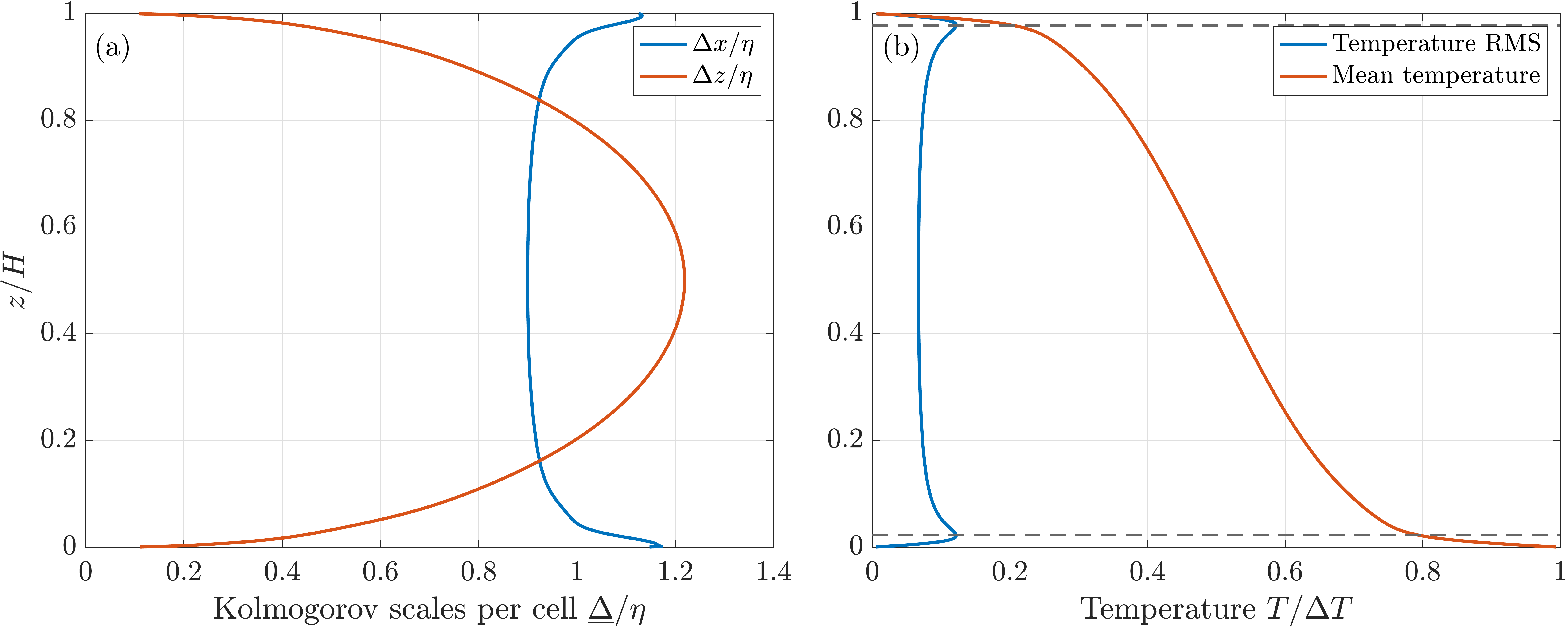}
    \caption{Number of Kolmogorov scales per cell (a) and temperature profiles (b) for the example case $Ra=10^7$ (and $\Ek=10^{-4}$, $Pr=1$). In (b), the dashed lines indicate the boundary layer edges based on the maximum of the RMS temperature.}
    \label{fig:validation}
\end{figure}

The different spatial resolutions that are used in this work are included in table~\ref{tab:input} of the main text. To validate these resolutions, we separately consider the bulk flow and the boundary layers. For the bulk resolution, we compare the grid spacing with the Kolmogorov length $\eta$ for the smallest kinematic features, and the Batchelor length $\eta_T$ for the smallest thermal features of the flow. Their respective definitions \citep{Monin1975} can be rewritten into convective units (using the free-fall velocity scale $U$ and length scale $H$) as
\begin{equation}\label{eq:kolmog_batchelor}
    \tilde{\eta}=\left(\frac{\Pran}{Ra}\right)^{3/8}\tilde{\epsilon}^{-1/4},\qquad \tilde{\eta}_T=\tilde{\eta}\Pran^{-1/2},
\end{equation}
where $\tilde{\epsilon}$ denotes the kinetic energy dissipation rate
\begin{equation}
    \tilde{\epsilon}=\sqrt{\frac{\Pran}{Ra}}\big|\tilde{\boldsymbol{\nabla}}\tilde{\boldsymbol{u}}\big|^2.
\end{equation}
In our work $\Pran=1$, so that the Batchelor length and Kolmogorov length coincide $\eta=\eta_T$. We can compare the Kolmogorov length to the local grid spacing $\underline{\Delta}=(\Delta x, \Delta y, \Delta z)$ in each dimension from a posteriori horizontal and temporal averages of the kinetic dissipation. We calculate the number of Kolmogorov lengths per cell in each direction $\underline{\Delta}/\eta$, as is shown in the example in figure~\ref{fig:validation}a.\footnote{Note that the grid is uniform in the horizontal direction, so we have $\Delta x /\eta = \Delta y /\eta$.} In all simulations, we ensure that we maintain $\underline{\Delta}/\eta<2$ over the entire vertical extent of the domain, which is well below the limit of $\underline{\Delta}/\eta<4$ that was empirically found to be acceptable by \citet{Verzicco2003}. Also for the ensemble of runs, where we use a coarser grid for computational feasibility (see table~\ref{tab:input}), we can adhere to $\underline{\Delta}/\eta<2$ throughout the full domain, owing the moderate $Ra=6\times10^6$.

To properly resolve the boundary layers at the top and bottom, we require that a sufficient number of grid cells reside within these boundary layers. Since this work uses stress-free boundary conditions for velocity, there is no formation of any kinematic boundary layers. For the thermal boundary layer, on the other hand, we adopt the definition of maximum (horizontally and temporally averaged) root mean squared (RMS) temperature \citep[e.g.][]{Julien2012}, see figure~\ref{fig:validation}b. We ensure that there are at least 10 grid cells within the thermal boundary layer for all simulations, which is also empirically coined sufficient by \citet{Verzicco2003}.

\vspace{5cm}

\bibliographystyle{jfm}
\bibliography{main}

\end{document}